# Experimental Demonstration of Real-time PDCP-RLC V-RAN Split Transmission over Fixed XGS-PON access

A. El Ankouri[(1),(2)], L. Anet Neto[(1)], A. Sanhaji[(1)], S. Barthomeuf[(1)], H. Le Bras[(1)], B. Le Guyader[(1)], A. Chagdali[(1)], M. Wang[(1)], N. Genay[(1)], K. Grzybowski[(1)], S. Durel[(1)], P. Chanclou[(1)]

[(1)] Orange Labs, 2 Avenue Pierre Marzin - 22300 Lannion, France, anas.elankouri@orange.com
[(2)] IMT Atlantique, 655 Avenue du Technopôle, 29200 Plouzané

**Abstract** In this work, we experimentally assess the transmission of a PDCP-RLC virtualised RAN split interface through a commercial XGS-PON system. We investigate the impacts of DBA on the uplink and packet jitter on the downlink.

**Introduction**

Mobile networks have greatly evolved from the couple of hundreds of kbit/s in 2G to the hundreds of Mbit/s in 4G. Right now, the 5G is on its way with expected bit-rates in the range of Gbits/s[1]. A great deal of efforts is being put in to allow the increase in capacity of the air interface while introducing the least possible changes to the legacy fibre infrastructures connecting the antenna sites to the central offices.

The bit-rate increase foreseen in 5G is fomenting the rise of interesting applications, amongst which providing fixed access connectivity to users in rural areas. In this context, Fixed Wireless Access (FWA)[2] is often presented as a competing technology to low bit-rate fixed access solutions such as Asymmetric Digital Subscriber Line (ADSL). In a nutshell, FWA allows providing broadband internet to house-holds through a mobile network such as 4G, as depicted in Fig. 1a, thus avoiding costly greenfield fibre deployments. The main challenge in this case is that fixed access prone services such as video streaming and online gaming impose higher constraints on the needed throughputs between the antenna sites and mobile evolved packet core (EPC) compared to standard mobile usages such as emailing and social media communication.

As opposed to this use-case, another interesting application can be considered with the 5G, this time for dense or ultra-dense areas. Indeed, the mobile traffic can be offloaded through available high capacity passive optical networks (PON), as seen in Fig. 1b. This would not only allow leveraging on already deployed fibre to the premise (FTTx) infrastructures but also alleviating the throughputs in the radio access network (RAN) while offering, through simple handover, seamless connectivity for the users on the transitions between typical mobile and fixed access contexts. This solution would also provide the unification of the wireless interfaces, from currently two (mobile and WiFi) to only one (5G), with higher capabilities in terms of air interference management. Moreover, large bandwidths could be targeted in the unlicensed or shared spectrum ranges[3]. In Fig. 1b, the customer premises equipment (CPE) embeds both fixed access and functional split based mobile functions. The RU+DU block carries layer 1 and lower layer 2 functionalities whereas a virtualized central unit (vCU) hosts higher layer 2 and layer 3.

We experimentally assess the transmission of 3GPP's option 2 interface[4] through a symmetrical 10G capable PON (XGS-PON)[5] access. This functional split is implemented on top of an Openstack virtualization environment together with the mobile evolved packet core (EPC). We investigate the impacts of dynamic bandwidth allocation (DBA) in the uplink transmissions and assess the robustness of the mobile interface to packet jitter introduced by an Ethernet traffic impairment engine that emulates degradations from the aggregation network.

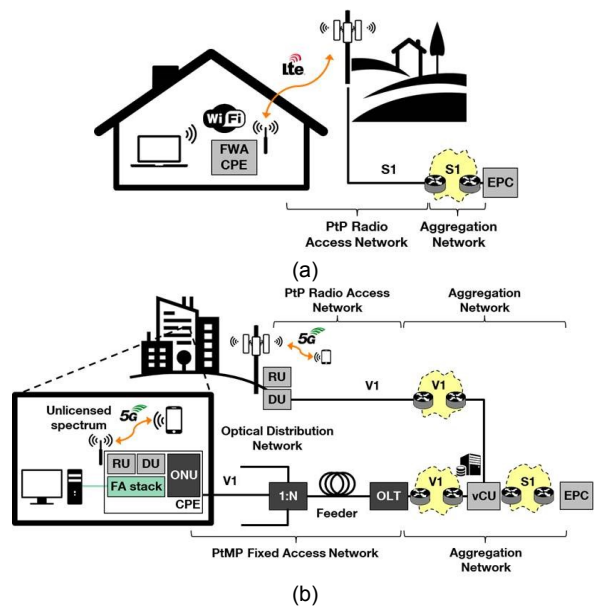

**Fig. 1:** (a) Rural and (b) dense area **FWA** scenarios.

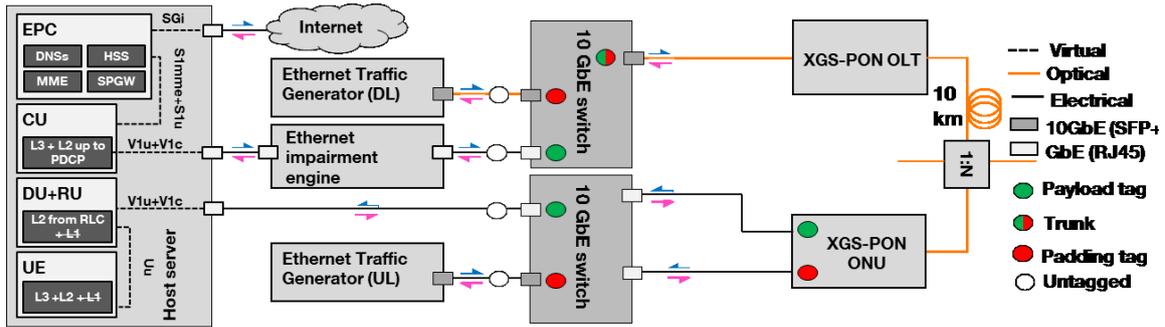

**Fig. 2:** Experimental setup

**Experimental setup**

Fig. 2 shows our experimental setup, which can be viewed as three distinct parts:

The RAN blocks run on a host server containing the LTE protocol stack and EPC. They are implemented on a CentOS/Openstack virtualization environment on provisioned virtual machines (VM). Each set of services physically separated in an actual mobile network are logically separated in our server. For instance, the EPC, containing a domain name system (DNS) server, mobility management entity (MME), serving and packet data network gateway (SPGW) and the home subscriber server (HSS) communicates with the CU via the S1 (backhaul) interface. The CU is connected to the DU via an Ethernet encapsulated V1 (3GPP PDCP-RLC split) interface which goes out of the server. It then goes through our access and (emulated) aggregation network segments and loops back to our server, in the DU+RU block. The CU hosts layer 3 and higher layer 2 functions whereas the DU+RU hosts lower layer 2 functionalities. Layer 1 functionalities of the RU and user equipment (UE) are abstracted without loss of generality since our main goal is to evaluate the transmission of a high layer split through an optical transmission system. The UE VM is also implemented and provisioned in the same server.

The aggregation network is emulated with an Ethernet Degradation Engine. It allows us to add packet latency and jitter to an incoming V1 interface. The mobile interface is then transported through a 10Gigabit Ethernet switch and mixed with a 8 Gb/s overloading Ethernet signal which emulates Fixed Access Network (FAN) traffic. The two streams are tagged with different VLAN tags by the switch and then transported together in the same port towards our Fixed Access Network in the downlink. An equivalent scenario is set up for the uplink but with different ports of the switch associated to the different tags so that the dynamic bandwidth allocation procedures can be individually applied to mobile and overloading traffic.

The FAN is composed by an WGS-PON optical line termination (OLT), and optical network unit (ONU), fixed attenuation emulating a splitter and 10 km of standard single mode fibre. Since PON operate under time division multiple access (TDMA) to allow sharing of the feeder fibre, a traffic container of type 3 (T-CONT 3) is associated to each ONU port. The DBA profile set on both ports allows for 150 Mb/s of assured traffic with the remainder of the capacity of the channel allocated on a best-effort basis. At the output of our ONU, we have two physically distinct streams transmitted over Ethernet. The streams are received by a second switch, which de-tags the streams before sending them back to the host server.

**Results and discussion**

We start our experimental assessment by investigating the impact of eventual degradations coming from the aggregation network. Fig. 3 shows the effect of the induced packet jitter to our downlink V1 interface, with user datagram protocol (UDP) packets 1200 bytes long. The packet jitter values correspond to the standard deviation of a normally distributed latency, with 2 ms mean value. Three sets of measures are shown, namely, without packet jitter, with 0.1 ms jitter and with 0.66 ms jitter. We can see that for a transmission without added jitter, the PER (Packet Error Rate) is relatively low from 20 to 100 Mbit/s. After that, we have a steep ascent up to 150 Mbit/s. We believe that such limitation is coming from the limited resources in our virtual machines. When adding packet jitter, the overall PER is higher and we observe a steep ascent that begins earlier with respect to the bitrate, with much stronger impact with 0.66 ms.

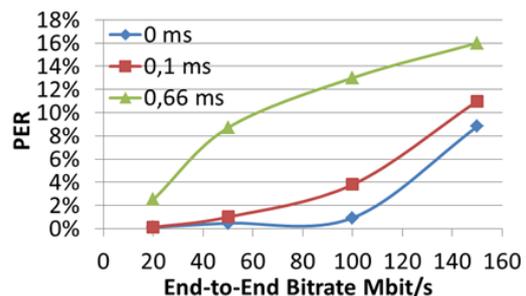

**Fig. 3:** PER variation with bit-rate for different induced packet jitter values.

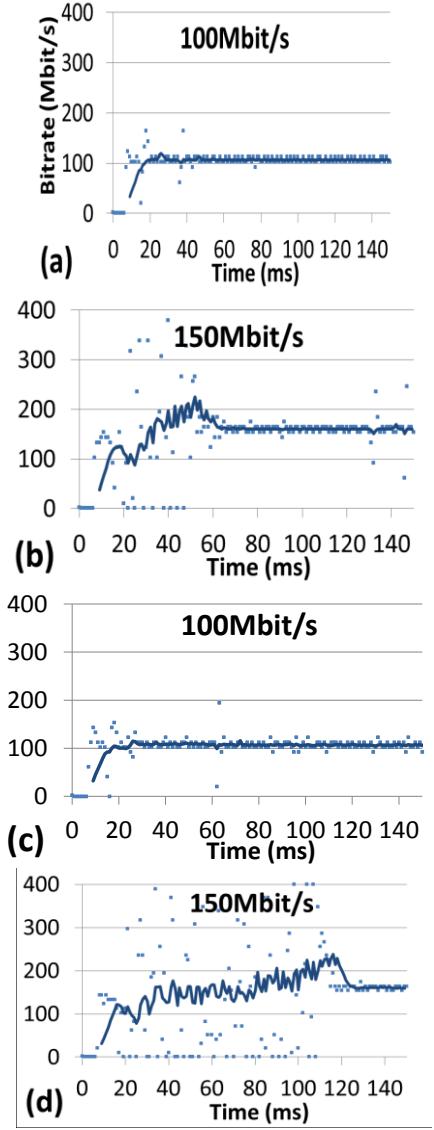

**Fig. 4:** Upstream bitrate evolution for point-to-point B2B (a and b) and uplink XGS-PON (c and d).

The second set of measurements consisted on investigating the impacts of DBA in the uplink by assessing the needed response times to provide full assignment of a specific bit-rate to the mobile stream. UDP packets are transmitted via iperf from the UE to the EPC while the overloading traffic is always set to a maximum useful bit-rate of 8.5 Gb/s. We then measured the evolution of the bitrate in the V1 interface of the CU, which is shown in Fig. 4. Two bit-rates are considered, one below the maximum assigned bit-rate (100 Mb/s) and another at the limited assigned capacity (150 Mb/s). Fig. 4 compares the evolution of the bit-rate with and an optical back-to-back (B2B) point-to-point scenario, where the server is looping back directly in itself and thus without TDMA (Fig. 4 a and b). Fig. 4c and d show the upstream bit-rate evolution profile for our XGS-PON system.

Comparing the two setups, we can see that the bandwidth is more rapidly allocated in the B2B case compared to the XGS-PON. For 150 Mbit/s, the B2B setup allocates the required bandwidth in 60 ms to the CU, whereas in the XGS-PON setup, a two-fold increase in the system response is observed, with 120 ms. We can also observe that without congestion (100 Mb/s bit-rates), the XGS-PON and point-to-point set-ups behave similarly, as expected.

| Round Trip Time (ms) | B2B | PON |
|---|---|---|
| min | 0,405 | 0,675 |
| average | 0,516 | 0,942 |
| max | 0,678 | 1,556 |
| std deviation | 0,047 | 0,154 |

**Tab. 1:** RTT comparison between B2B and 10 km XGS-PON

Finally, Tab. 1 shows the round trip time (RTT) latencies added by our XGS-PON system over 10 km compared to the point-to-point back-to-back scenario. It can be observed that the XGS-PON not only increases the mean latency but also increases the packet jitter by a factor of 10, from 0.047 ms to 0.154 ms, which could impact the mobile transmission, depending on the service.

**Conclusions**

In this work, we demonstrated the feasibility of FWA for high dense area by transporting a mobile traffic through an emulated aggregation network and a fixed access network using XGS-PON technology. We assessed the impacts of introduced packet jitter to the downlink as well as the degradations induced by DBA profiles with limited assured bit-rates.

**Acknowledgements**

This work was supported by the European H2020-ICT-2016-2 project 5G-PHOS.